# Summarization in Semantic Based Service Discovery in Dynamic IoT-Edge Networks

Hessam Moeini, I-Ling Yen, Farokh Bastani

*Erik Jonsson School of Engineering & Computer Science - Department of Computer Science*
*University of Texas at Dallas*
Email: {moeini, ilyen, bastani}@utdallas.edu

**Abstract**— In the last decade, many semantic-based routing protocols had been designed for peer-to-peer systems. However, they are not suitable for IoT systems, mainly due to their high demands in memory and computing power which are not available in many IoT devices. In this paper, we develop a semantic-based routing protocol for dynamic IoT systems to facilitate dynamic IoT capability discovery and composition. Our protocol is a fully decentralized routing protocol. To reduce the space requirement for routing, each node maintains a summarized routing table. We design an ontology-based summarization algorithm to smartly group similar capabilities in the routing tables and support adaptive routing table compression. We also design an ontology coding scheme to code keywords used in the routing tables and query messages. To complete the summarization scheme, we consider the metrics for choosing the summarization candidates in an overflowing routing table. Some of these metrics are novel and are difficult to measure, such as coverage and stability. Our solutions significantly reduce the routing table size, ensuring that the routing table size can be bounded by the available memory of the IoT devices, while supporting efficient IoT capability lookup.

Experimental results show that our approach can yield significantly lower network traffic and memory requirement for IoT capability lookup when compared with existing semantic based routing algorithms including a centralized solution, a DHT-based approach, a controlled flooding scheme, and a cache-based solution.

---------- ◆ ----------

## 1 INTRODUCTION

IN recent years, IoT (Internet-of-things) technologies have advanced significantly. However, many existing IoT systems are statically built. In these systems, the specific IoT devices and control and management software are statically selected and configured at the design time to achieve some predefined tasks and to handle some anticipated events. This type of systems has a similar nature as the conventional embedded systems, except that the constituent components (devices and software) are distributed in a wider area. We attempt to expand the scope of IoT by considering dynamic IoT systems to address dynamically arising tasks. To make the best use of a large number of available IoT devices, we should be able to discover and compose IoT capabilities in the IoT network to respond to dynamically arising tasks.

An important requirement for dynamic composition of IoT systems is to dynamically discover the IoT services based on their functionalities, i.e., semantic based routing. There have been a lot of research works in semantic based routing [1 - 4], but they have some limitations when applied to IoT systems. Centralized or supernode based schemes [1, 2] are not scalable for widely distributed IoT systems, especially when we consider mobile IoT devices, which may cause frequent updates and result in significant communication overhead.

Decentralized semantic-based routing protocols include structured and unstructured schemes. Structured semantic routing solutions are mostly DHT (distributed hash table) based [3, 4], which hashes the resources or services to specific servers. They are very effective for digital objects or software services but are not applicable to IoT devices because IoT devices cannot simply be moved to the hashed locations. Unstructured routing approaches include table driven and information caching based solutions. Table driven routing protocols maintain a routing table in each node via advertisements [5]. Information caching schemes maintain a similar routing table in the cache of each node, but only via past routing results without advertisements [6, 7]. These schemes may result in a large routing table (or cache) size, potentially having one entry for every capability in the system.

Since existing semantic based routing algorithms are mostly designed for peer-to-peer systems, they do not need to consider the memory size limitations. But memory space limit is a major concern for IoT nodes. An easy solution to the memory constraint is to delete some routing information when the routing table size exceeds a given limit. But by doing so, some potentially useful routing information may be lost. In IP based routing protocols, similar routing table size problem is resolved by dominant address aggregation and route summarization, or Classless Inter-Domain Routing (CIDR) [8], where IP addresses with a common prefix are aggregated in the routing tables. However, in semantic-based routing, how to achieve a similar and effective summarization when keywords are used as the routing table indices? In this paper, we develop a sematnic-based routing algorithm with a novel summarization solution. Thus, the algorithm can observe routing table size constraint while retaining useful routing information. (Note that for convenience, we use routing table to refer to the routing information maintained in both table-driven and information cachcing approaches).

**How to perform routing table summarization in semantic based routing?** The works addressing the summarization issue for semantic based routing is limited, [5, 9]. In fact, none of them provides a complrehensive solution. CBCB [5] considers summarization, but only for overlapping numerical ranges. It also supports logical composition (conjunctive) of keywords, but it does not help with routing table size reduction.

In our solution, we organize the capabilities of the IoT devices in an **ontology tree**. In the ontology tree, similar capabilities have a common ancestor. Leaf nodes in this tree are the actual capabilities of the IoT devices and the internal tree nodes are the capability categories for their subtrees. Consider capabilities $a$, $b$, and $c$, get summarized into $d$ in the ontology tree of IoT capabilities. If a node $n$ in the IoT network has $b$, $c$, $d$ capabilities from a neighbor $y$ in its routing table, then, $x$ can replace $a$, $b$, and $c$ by $d$ for neighbor $y$. This summarization technique can be used to effectively control the routing table size to fit the memory constraint of each node. Based on the specific space limit given by a node for its routing table, summarization can be done adaptively, i.e., perform more aggressive summarization, such as summarizing $b$ and $c$ into $d$ or summarizing capabilities into a grandparent or even higher layer of a node in the tree, when space becomes tighter. An aggressive summarization may cause misleading routing for lookup queries, but it may still be better than throwing away the information completely in cache based solutions.

**How to know which capabilities can be summarized into which categories?** A problem arises in the summarization approach. How would a node in an IoT network know that capabilities $a$, $b$, and $c$ can be summarized into $d$. Also, when a query requesting for $a$ get routed to $n$ and $n$ has capability $d$ in its routing table, how would $n$ know that $d$ should be used for routing $a$. Obviously, each routing node needs to maintain the whole ontology tree in order to perform summarization and corresponding routing. But this will incur a high overhead in memory usage, defeating the purpose of summarization. Also, traversing the ontology tree to locate the relevant capabilities for potential summarization or routing can put significant computing burden on resource-constraint IoT nodes.

We develop an **ontology coding** technique to support summarization without needing to store the ontology tree on each node. Instead of using the keywords to represent the capabilities, the corresponding ontology codes are used for indexing the routing table and for respenting the lookup entries in the queries. The code is informative and can be used to determine whether the capabilities can be summarized and how to summarize them without the ontology tree.

**How to choose the entries for summarization?** We assume that each node in the IoT network gives a memory bound for its routing table. When the routing table size exceeds this limit, summarization is performed. But which entries should be summarized, and which ones should remain as is? We introduce four metrics for determining the summarization candidates. The "hop distance" metric shows how close a capability provider is to the querier node. "Usage" has the same concept as conventional caching schemes, and it keeps track of the usage frequency and recency for each capability in the routing table. The "coverage" metric considers how comprehensively a summarized ontology capability is covered in a neighborhood, if it is summarized without all capabilities in its subtree being covered. "Stability" helps a node to understand how long the received routing information will last due to its providers' mobility patterns. We integrate these metrics to determine the best choices of summarization candidates.

Based on the ontology-based summarization technique, we design a fully decentralized semantic-based routing protocol for dynamic IoT networks. The major contributions of this paper are discussed in the following.

- In our approach, instead of throwing away routing data, we introduce a novel *ontology*-based approach to summarize them to retain the potentially useful information while significantly reduce their space requirements. To the best of our knowledge, we are the first to design a capability summarization technique to enable summarization of keywords representing the capabilities and achieve comprehensive summarization in semantic-based (or content-based or information-centric) routing.

- In most of the existing semantic based routing algorithms, keywords are directly used in the routing table. We introduce an ontology coding scheme to represent the keywords and capture its relative relations in the ontology tree in order to support summarization without needing to keep the ontology tree on each node in the IoT network. Our coding scheme is novel and can help greatly reduce the routing table size as well as query message size.

- To complete the summarization scheme, we consider the metrics for choosing the summarization candidates in an overflowing routing table. Some of these metrics are novel and are difficult to measure, such as coverage and stability. We design methods to estimate them to spport better summarization candidate selection.

- We conduct thorough experimental studies to evaluate the effectiveness and efficiency of our protocol. Our approach is compared with several representative semantic-based routing solutions in the literature, including a centralized solution, a DHT-based approach, a controlled flooding scheme, and a cache-based solution. Experimental results show that our approach consumes less memory space and imposes lower volume of network traffic compared to all other approaches.

The rest of the paper is organized as follows. In Section 2, we survey the existing works. The problem specification is given in Section 3. Section 4 discusses ontology based summarization approach and the ontology coding scheme. Section 5 introduces the metrics required to calculate the utility values for capabilities in the routing table and enabling the intelligent capability grouping in the summarization procedure. Section 6 introduces the routing algorithm that integrates the techniques discussed earlier. Experimental study and performance comparisons are discussed in Section 7. Finally, Section 8 states the conclusion of the paper, discusses some limitations of our protocol, and outlines future research directions to address these limitations.

## 2 EXISTING WORKS

**Existing Semantic-based Routing Protocols.** Some semantic-based routing protocols are centralized or hierarchical [1, 2]. They require the IoT devices to register their capabilities at the registry node(s) and keep them updated upon changes. These schemes may incur a high overhead in more dynamic IoT networks (with higher mobility or high rates of nodes joining and leaving) because a large number of update messages may need to traverse through a long distance to get to the registry. Also, locality is another concern for these schemes. A local query requesting for a nearby IoT capability may need to go far away to the central registry and back, which may impose additional network traffic and increase lookup latency.

Decentralized semantic-based routing protocols can help overcome the problems discussed above. Existing semantic-based routing protocols for peer-to-peer (p2p) systems include structured and unstructured solutions. Structured routing protocols are mostly DHT (distributed hash table) based [3, 4]. They are designed for documents and are not suitable for IoT devices because we cannot move IoT nodes to where they are hashed to. Using the node at the hashed location as a pointer can solve the problem, but may yield a high network traffic and latency, which may have more severe performance problems than the centralized solutions.

Unstructured routing protocols are mainly flooding, and information caching based. Cache-based approaches [6, 7] cache the information of previously discovered resources to guide routing of future queires. These schemes may result in a large cache size, potentially having one entry for every capability in the system. GSD [7] adds a hop limit to confine the cache size. It categorizes capabilities based on their similarities that has some similarity to the IoT ontology we consider but without using it for the summarization. Each capability can be a member of different groups. Each node advertises its capability(es) to the limited local region. Advertisement message also piggybacks a set of capability's groups which the routing node has received previously. Therefore, each node has a knowledge of its own local region, as well as available groups in further distances. A lookup query for capability $x$ includes $x$ itself and the groups it belongs to. During lookup, node will first look for $x$ in its cache and forwards the query to all the neighbors if it exists in the local region. If $x$ is not available in the cache (i.e. $x$ is not in the local region), node checks the stored groups in the cache and forwards the query to a suitable next hop based on the $x$'s groups in the query message. If such a group does not exist in a routing node, it will do flooding again. This approach is not efficient in terms of network traffic and still needs to do flooding whenever the capability is in the local region. Also, nodes in this approach cache a lot of information (such as capability description, capability groups, etc.) which is wasting memory space specially because they also use plaintext keywords to do that. In GSD, cache entries and messages are indexed by the nodes' IDs and have different length which results in more complicated processing in order to do matchmaking process. These specefications (using plaintext keywords and the indexing scheme) also resulted in no cache summarization method possible in this approach.

ICN (Information Centric Network) can also be considered as a semantic based routing technique. Several routing approaches have been considered in ICN, including *name resolution* and *name-based* routing [10 - 12]. The name resolution approach requires a name resolution service (NRS) in the system to map named capabilities to the actual resources. The NRS resolution scheme is similar to the mechanism used in the Internet domain name server (DNS). It considers large-scale networks, but may incur a high network traffic and latency. Name-based routing solutons [10] directly route the queries to the IoT nodes using resolution handlers (RHs). These solutions are very similar to the information caching solutions and require storing all the routing information in the caches to help query routing. We need to address memory space limitation in resource-constraint IoT devices.

**Existing Summarization Methods.** Existing semantic based routing protocols [5, 9] are mainly designed for peer to peer systems where each node has sufficient resources for handling routing. In resource-constraint IoT devices, we need to consider space limits for routing information storage (in table based or information caching schemes). Though the space problem in these schemes can be easily resolved by using traditional cache replacement protocols, yet such solutions may cause the eviction of useful routing information. On the other hand, the summarization technique can best retain useful routing information upon tight space situaitons. In IP based routing, summarization, such as dominant address aggregation or Classless Inter-Domain Routing (CIDR), is a common practice where IP addresses with a common prefix are aggregated in the routing tables. The challenge is: how to achieve effective summarization when keywords are used to index the routing tables? Though no existing semantic based routing protocols consider summarization, there are some partial solutions [5, 8]. CBCB [5] considers and/or connectives of capability's attributes in the routing table in order to guide discovery query routing correctly. But putting many combinations of multiple keywords into the routing table may make the routing table size grow unreasonably. CBCB introduces a *covering* concept which helps nodes to ignore more detailed capabilities if they already have a more general one covering that received information. Covering scheme is similar to the summarization technique, to reduce routing table size but it can only work for numerical data and logical relationships, not plaintext keywords. Moreover, covering only can support simple scenarios and is not efficient enough. For example, consider a node $n$ has a predicate $p_1$ for an attribute $x$ with the value between 2 and 10. Assume $n$ receives another predicate $p_2$ from the same interface with the attribute $x$ and values greater than 10. An efficient summarization is when $n$ combines these two entries and stores predicate $p_3$ for $x$ values greater than 2. However, node $n$ in CBCB cannot handle this scenario because none of the predicate $p_1$ and $p_2$ covers the other one.

Bloom filter has been used in some semantic based routing scenarios [13 - 15]. Bloom filter by its nature summarizes the capability keywords in the filter by hashing. It achieves space as well as lookup efficiency. It is suitable for hierarchical routing architecture where multiple "registry" servers exchange their entries via Bloom filters. However,

Bloom filter is only space efficient when it carries relatively dense information. For example, consider a system with 1000 different IoT capabilities. To confine the false positives rate within 0.01, the Bloom filter should be of size 5.85 KB for 1000 keywords. Assume that each node only gets 100 capabilities. Also assume that each keyword is of 12 characters (12 bytes) in average. If we directly use keywords, then each routing table would be of 1.2 KB, while the Bloom filter requires 5.85 KB.

## 3 PROBLEM SPECEFICATION

We consider a large number of IoT devices distributed over many wireless networks which are interconnected via Internet or cellular networks. Each node in the network has one or more IoT capabilities or is a routing node. We assume that each node in the network can discover its neighbors using some existing techniques, such as the mesh-based robust topology discovery algorithm [16] or the topology construction phase in RPL. If an IoT device does not have sufficient resources to support routing, we assume that one neighbor node will be responsible for its routing.

Generally, each IoT device has a main functionality and we refer to this as its "capability". An IoT device may have multiple "capabilities". In semantic and service computing, keywords can be used to define the capabilities of IoT devices to facilitate search and composition of capabilities to realize specific tasks. Let $cpbs(n)$ denote the set of capabilities of a node $n$. Multiple nodes, e.g., $n_1$ and $n_2$, in the IoT network may have the same capability, $cpbs(n_1) = cpbs(n_2)$. For example, both picture camera and video camera have the capability of capturing still images. Video camera has an additional capability of capturing videos.

Instead of flooding or random routing, we consider that each node maintains a routing table (or information cache) to facilitate effective routing. Since many nodes in an IoT network may have limited memory spaces, their routing table size should be bounded. Let $RT(n)$ denote the routing table of node $n$ and $RTB(n)$ the size limit for $RT(n)$.

An IoT service discovery query $Q$ can be issued by any node in the network. $Q$ is defined by $Q(cpb, sn, ttl)$, where $Q.cpb$ is the requested capability, $Q.sn$ is the source node issuing $Q$, and $Q.ttl$ is to limit the number of hops in lookup routing for $Q$. The goal of the semantic-based routing protocol is to find one IoT device (node) that provides the capability $Q.cpb$, i.e., $Q.cpb \in cpbs(dn)$, where $dn$ is the node that can provide the desired IoT service described by $Q.cpb$. Note that there may be more than one node providing $Q.cpb$ in the network. Service discovery algorithms should be able to find single or multiple service provider(s). In this paper we consider that querier is only asking for one service provider. In case multiple service providers are needed, query can easily be modified to represent that. Correspondingly, a node forwards the query in multiple directions where multiple service providers exist and have been advertised from. For example, node $n$ receiving a query $q$ asking for ten IoT devices with the capability $x$ will search its routing table and forwards the query to one or multiple neighbors (if available) who have advertised $x$ for different providers. $n$ then updates the query with a new desired number of providers to prevent unnecessary propagation in different directions. Our protocol does its best effort to find the desired number of requested service provider if they exist in the IoT network. However, multiple neighbors may have advertised the capability for the same provider, and this can result in a problem. If needed, this problem can be handled by adding the providers' ID to the routing table.

## 4 ONTOLOGY BASED SUMMARIZATION

We use an ontology to enable capability summarization in discovery query routing. An example ontology is shown in Fig. 1 (a). It shows various classes of "Imaging" devices, including cameras, medical cameras, and radars. Under camera, there are "Picture" and "Video" cameras which are further divided into more specific capabilities based on their resolutions and mobility. Generally, each IoT device can be categorized into one of the leaf nodes in the ontology. For convenience, we lable the ontology nodes, such as T11 for "Mobile-LowResolution-DPC (digital picture camera)", T12 for "Fixed-LowResolution-DPC", etc.

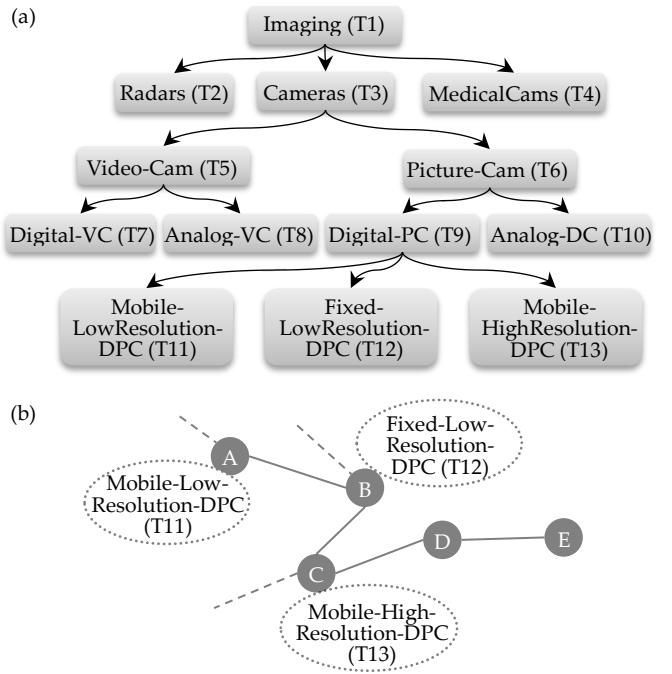

Fig. 1. (a) A sample Ontology (b) A sample network

Consider a sample network as shown in Fig. 1(b). It consists of nodes A, B, C, D, and E. Nodes A, B, and C have capabilities T11, T12, and T13, respectively. A, B, and C advertise their capabilities which progressively reaches D. D summarizes the capabilities in its routing table by replacing T11, T12, and T13 by T9 for neighbor C. If C's routing table size exceeds its bound, i.e., $size(RT(C)) > RTB(C)$, C may aggressively summarize T11 and T12 into T9 in its routing table for neighbor B.

To achieve summarization, routing nodes may need to store the entire ontology for reference. Consider the previous example. Without the ontology, how would D summarize T11, T12, and T13 into T9. Also, when D receives a query to lookup T12, how would it know that T9 in its routing table should be used. On the other hand, it may not be feasible for many routing nodes to store the entire ontology. We design an ontology coding scheme, where each capability in the ontology is represented by a code. The routing table of each node is indexed by the ontology code instead of the capability keyword, and the query uses the code for lookup. Summarization and routing can be done based solely on the code without referencing the original ontology tree. We call the code for each ontology node its ONID (Ontology ID) and ONIDs should satisfy:

- **Uniqueness.** As with any coding schemes, the code for each ontology node should be unique.
- **Informative.** From the ontology code, a routing node should be able to recognize the sibling and parent-child relationships in the original ontology tree so that summarization and query routing can be performed. Also, from the code of a child capability, a routing node should be able to construct the code for the parent capability in the ontology (when we need to replace the child capabilities by the parent capability).

### 4.1 Ontology Coding Scheme (OCS)

Let $ONID(on)$ denote the ontology id of an ontology node $on$. $ONID(on)$ consists of two bit vectors, including the "ID" bit vector and the "SP" bit vector. The ID vector specifies a code for each ontology node. It is an aggregation of codes level by level from root to the node in the ontology tree. The SP vector specifies the "starting position" of each level of code. Let $ONID(on).ID$ and $ONID(on).SP$ denote the ID and SP bit vectors of an ontology node $on$. $ONID(on).ID$ includes the parent code, $ONID(on).PID$, and the sibling code, $ONID(on).SibID$. $ONID(on).PID$ is essentially the ID vector of on's parent. The sibling code is a unique code among the siblings of node n. More formally, we have

$$ONID(on).ID = ONID(on).PID \bullet ONID(on).SibID$$
$$ONID(on).PID = ONID(parent(on)).ID$$
$$ONID(on).SibID = pos(n, chlist(parent(on)))$$

Here, function $parent(on)$ returns the parent node of on in the ontology, $chlist(on)$ returns the list of child nodes of node $on$, and $pos(on, l)$ returns the position of on in list l (assume that on is an element of l). To uniquely define the sibling code $ONID(on).SibID$, its code length should be $\|ONID(on).SibID\| = \lceil \log_2 \|chlist(parent(on))\| \rceil$

Fig. 2 shows an example of coded ontology nodes. The root ontology node has an assigned code "0". The root has five children. For all the child nodes $i, 1 \leq i \leq 5$, their $ONID(i).PID$ should be "0", $\|ONID(i).SibID\|$ should be 3 bits, and $ONID(i).SibID$ should be "000", "001", "010", "011" and "100". As shown in the figure, the same coding scheme is applied to the three child nodes of "0001".

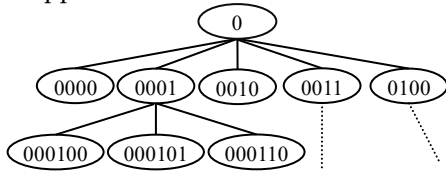

Fig. 2. $ONID(on).ID$

As can be seen, the coding scheme defined above will result in different code length for each ontology node. If we simply pad the code, then the ID for each node may not be unique. More importantly, from $ONID(on).ID$, we cannot decompose the code to recognize $ONID(i).PID$ and $ONID(i).SibID$ which is essential for identifying the relations between nodes.

We use the other component of ONID, the SP vector, $ONID(on).SP$, to resolve the problem. $ONID(on).SP$ specifies the "starting position" of each level of code in $ONID(on).ID$. Similar to $ONID(on).ID$, $ONID(on).SP$ includes the parent SP, $ONID(n).PSP$ and the sibling SP $ONID(on).SibSP$. To indicate the starting position, each piece of SP code has its leftmost bit set to 1 and the remaining bits are 0. Formally, we have

$$ONID(on).SP = ONID(on).PSP \bullet ONID(on).SibSP$$
$$ONID(on).PSP = ONID(parent(on)).SP$$
$$\|ONID(on).SibSP\| = \lceil \log_2 \|chlist(parent(on))\| \rceil$$
$$ONID(on).SibSP[i] = \begin{cases} 0, \text{if } i = 0 \\ 1, \text{otherwise} \end{cases}$$

Fig. 3 shows the corresponding SP vectors for the sample ontology tree given in Fig. 2.

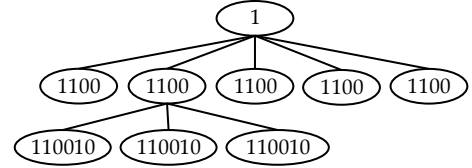

Fig. 3. $ONID(on).SP$ vector coding

Now we can pad $ONID(on).ID$ and $ONID(on).SP$ with 0's for all ontology nodes to the same length. The ONID for each node is the concatenation of $ONID(on).ID$ and $ONID(on).SP$.

### 4.2 OCS Properties

Uniqueness of the ONID can be shown by contradiction. Assume that two different ontology nodes $on_1$ and $on_2$ have the same ONID $onid$. This implies

$$ONID(on_1).ID = ONID(on_2).ID \quad (1)$$
$$ONID(on_1).SP = ONID(on_2).SP \quad (2)$$

The ID vector of node $on$ contains the code pieces (SibID) of nodes from root to $on$ and the staring positions embedded in the SP vector can be used to divide the ID vector into pieces (each piece is a SibID). Let $\{sid_1^1, sid_2^1, \ldots, sid_{m_1}^1\}$ and $\{sid_1^2, sid_2^2, \ldots, sid_{m_2}^2\}$ denote the ID pieces obtained by such division for nodes $on_1$ and $on_2$, respectively. (1) and (2) imply that the same SP vector is used for dividing the same ID vector and, hence, we have $m_1 = m_2$ and $sid_i^1 = sid_i^2$, for all $i, 1 \leq i \leq m_1$. $sid_1^1$ and $sid_1^2$ ($sid_1^1 = sid_1^2$) have to be the ID vector of the root node. $sid_2^1 = sid_2^2$ implies that they refer to the same position in the children list of the same parent, i.e., the same node. Repeatedly, we can infer that $sid_{m_1}^1$ and $sid_{m_2}^2$ refer to the same position in the children list of the same parent, i.e., the same node. This contradicts the assumption that $on_1$ and $on_2$ are two different ontology nodes. Thus, two different nodes will not have the same ONID and our ONID guarantees uniqueness.

OCS provides informative codes for each capability in the ontology tree. Parent's ONID of each node in the ontology tree can be easily achived just based on its own ONID have knowledge of its different bit vectors. Another useful information that can be achieved from nodes' ontology IDs is to find the common ancestor of two given nodes in the ontology tree and their distance from each other. This information helps the routing table summarization procedure by letting network nodes to be able to list similar capabilities and replace them by their common ancestor in the routing table.

## 5 SUMMARIZATION CANDIDATE SELCTION

When the routing table reaches its space limit, we need to select the best entries to summarize. We introduce four metrics that should be considered in summarization candidate selection, including ontology coverage ($oc$), stability ($stb$), hop distance ($hop$) and historical usage ($ug$) and they are discussed in the following Subsections.

### 5.1 Ontology Coverage Metric

Consider a capability $rst$ which is the root of a subtree in the ontology tree. Let $ost$ (ontology subtree) be the set of all leaf nodes in $rst$'s subtree. Assume that node $n$ tries to summarize a set of capabilities in its routing table (let $lsc$ be the list of nodes for summarization) into $rst$. If $lsc = ost$ (all nodes in $ost$ subtree exist in the $lsc$), then $rst$'s coverage is 100% and the summeriztion is perfect. But if $lsc \neq ost$ (in fact $lsc \subset ost$) and assume that $lon$ is a leaf ontology node of $ost$ and it is not in $lsc$, then during query routing, $n$ will forward the query looking for $lon$ incorrectly. As can be seen, the coverage of $rst$ for summarization is a very important metric. If $rst$ is summarized with a low coverage, then there will be more misleading forwarding in routing.

$rst$'s ontology coverage, $rst.oc$, should consider the number of capabilities being covered by $rst$. Note that $rst$ replaces the capabilities in $lsc$ but does not just represent these nodes and it also represents all the other capabilities in $ost$ (in fact, the capabilities of all its descendants). Therefore, $rst.oc$ also depends on how well the capabilities in $ost$ are covered by the capabilities in the $lsc$ list. The coverage should consider both the number of capabilities being covered and the $oc$ values of the covered capabilities. The new $oc$ for $rst$ can be defined as:

$$rst.oc = \left\lceil \frac{\sum_{cpb \in lsc \wedge cpb \neq rst} cpb.oc}{\|ost\|} \right\rceil. \qquad (3)$$

$oc$ value for leaf nodes in the ontology is considered to be 1. $\|ost\|$ is the size of $ost$, i.e., number of capabilities in $ost$. However, it will be very expensive to store the ontology tree structure in each resource-constrained IoT node. Thus, it is not easy to obtain $\|ost\|$. Instead, we estimate $\|ost\|$ from the ontology code and the properties of the original ontology tree. Consider a node $on$ in the ontology tree. Recall that $\|chlist(on)\|$ is the degree of $on$ (i.e., the number of child nodes of $on$) and $\lceil\log_2\|chlist(on)\|\rceil$ is the number of bits used to code $on$'s children. We define sparseness of $on$ as $\|chlist(on)\|/2^{\lceil\log_2\|chlist(on)\|\rceil}$. If $on$ has 6 children, then it will require 3 bits to code its children and its sparseness is 6/8. Let $deg$ and $sparse$ denote the average degree and sparseness of the original ontology tree.

Consider $TL$ (tree level) as the height of $ost$, the subtree of the original ontology with node $rst$ as its root. If $TL = 1$, then we only need to estimate the number of children of $rst$. First, we find the node $mch$, which has the highest sibling ID, $sibID$, among the nodes in $lsc$ list, i.e.

$$ONID(mch).SibID = \max_{ch \in lsc \wedge ch \neq rst}(ONID(ch).SibID)$$

The maximal number of children nodes $rst$ may have is $2^{\lceil\log_2\|ONID(mch).SibID\|\rceil}$. With the consideration of potential sparseness of the subtree, we can estimate $\|ost\|$ as

$$\|ost\| = \max(ONID(mch).SibID,$$
$$\lceil 2^{\|ONID(mch).SibID\|} * sparse \rceil) \qquad (4)$$

If $TL = 2$, then we first identify the node $mch$, which is an immediate child of $rst$ and has the highest $sibID$ among its siblings in $lsc$ list, i.e.

$$ONID(mch).SibID = \max_{ch \in lsc \wedge parent(ch)=rst}(ONID(ch).Sib + ID)$$

Similar to (4), we decide the size of the first level subtree in $ost$, namely, $ost$-1,

$$\|ost\text{-}1\| = \max(ONID(mch).SibID,$$
$$\lceil 2^{\|ONID(mch).SibID\|} * sparse \rceil)$$

Next, we identify the grandchildren of $acpt$ and construct a maximal grandchild list $mgch$-$list$. If $mgch\_i$ is in $mgch$-$list$, then $mgch\_i$ is a grandchild of $acpt$ and $mgch\_i$ has the highest $sibID$ among its siblings, i.e.,

$$ONID(mgch_i).SibID = \max_{gch \in lsc \wedge parent(gch)=parent(mgch_i)}(ONID(gch).SibID)$$

For each node in $mgch$-$list$, we can compute the second level subtree size the same way as (4). But there may be nodes that are $rst$'s immediate child but are not in $lsc$ (there are $\|ost\text{-}1\| - \|mgch\text{-}list\|$ such nodes). For these nodes, we estimate their number of children to be $deg$. Thus the estimated total number of nodes in all the second level subtrees of $ost$, namely, $ost$-2, can be

$$\|ost\text{-}2\| = $$
$$\sum_{mgch_i \in mgch\text{-}list} \max(ONID(mgch_i).SibID,$$
$$\lceil 2^{\|ONID(mgch_i).SibID\|} * sparse \rceil)$$
$$+(\|ost\text{-}1\| - \|mgch\text{-}list\|) * degree.$$

Finally, the total estimated $ost$ size is

$$\|ost\| = \|ost\text{-}1\| + \|ost\text{-}2\|.$$

When $TL > 2$, $\|ost\|$ can be computed recursively similar to the $\|ost\|$ computation for $TL = 2$. For $TL = l$, we have

$$\|ost\| = \sum_{i=1}^{l}\|ost\text{-}i\|, \text{ and}$$
$$\|ost\text{-}i\| = $$
$$\sum_{m_i \in m\text{-}list} \max(ONID(m_i).SibID,$$
$$\lceil 2^{\|ONID(m_i).SibID\|} * sparse \rceil)$$
$$+(\|ost\text{-}(i\text{-}1)\| - \|m\text{-}list\|) * degree.$$

Here, $m$-$list$ is the list of capabilities in $lsc$ which are at the $i$-th level of $ost$.

## 5.2 Hop Distance Metric

Hop distance is a natural metric to determine the importance of a routing table entry. Capabilities that are further away are likely to be maintained redundantly among neighbors. If a capability *cpb* is far from a node *n*, then summarizing *cpb* and other capabilities into its ancestor may have less impact in queries routing. Along the route, more detailed information about *cpt* will be available.

*hop* shows how close *cpb* is from *n* considering neighboring node *j*. *n* needs to update the *hop* for *cpb* before storing *cpb* to its routing table to factor in the extra distance incurred. Due to the extra distance to the "source" of the capability, the new *hop* is incremented by one. We consider *hop* = 0 when node is advertising its own capability to the neighbor. Lower node-to-source distance value, higher utility, and higher utility shows more useful capability information in advertisement message and routing table.

## 5.3 Historical Usage Metric

In caching, historical usage information for each entry, such as the latest access time and the access frequency, are always used to decide the importance of an entry. Here, we use a weighted sum to aggregate the two factors and a power sum to process the historical usage frequency. Capabilities that are not used frequently or their last access times are long time back ago may either be removed from the routing table or summarized more aggressively.

The usage attribute computation for capability $c_i$ is defined as follows.

$$c_i.ug(t) = \alpha(freq(c_i,t)) + (1-\alpha)(c_i.ug(t-1)). \quad (5)$$

$freq(c_i,t)$ is the frequency of references to $c_i$ during time period $t$ and $c_i.ug(t)$ is a calculated value to show the usage of capability $c_i$ over the time period $t$. The higher $c_i.ug$ is, the more useful $c_i$ is.

When a service capability $c_i$ is newly propagated to $n$, it will not have historical access information on node $n$ and $freq(c_i,t)$ will be 0. To make sure that $c_i$ entry can be considered fairly during cache entry selection, we need to provide an initial $c_i.ug$ for a capability $c_i$ without historical access data. Assume that this capability entry $c_i$ is offered by node $n_x$. We require $n_x$ to send its own usage data for $c_i$ together with capability $c_i$ during the update message propagation and node $n$ will use it as $c_i$'s usage value.

If $c_i$ is an internal node in the ontology tree and is being considered newly for caching, then its usage data $c_i.ug(t)$ will have the same problem as above. In this case, we initialize $c_i.ug(t)$ to the average of its descendants' usage values.

## 5.4 Stability Metric

IoT networks include fixed, stable nodes, but quite a lot of IoT devices are mobile, which results in constantly changing IoT network topology. Summarization can reduce the routing table size, but node can mix high reliable information with low reliable ones received from unstable nodes (high-speed nodes) in the routing table and replace them with a summarized information. This can cause query routing misleading when nodes supporting low reliable information leave the neighborhood rapidly. A capability that is currently in range but will be out of range soon is not worth to be stored and participate in summarization.

The "stability" refers to the stability of the communication link between a node $n$ and a capability $c_i$ in $RT(n)$ (or the actual offering node of $c_i$). It is defined by how long the node which offers capability $c_i$ will stay within the communication range of node $n$ considering current routing path.

Consider $stb(n, c_i.n_j)$ as the stability value which node $n$ will have relatively to its neighbor $c_i.n_j$. Nodes will predict their mobilities in some future time intervals and will let other neighbors know that by the update messages. By doing comparison between the received information of $n_j$'s mobility and its own predicted mobility, node $n$ estimates the stability of $n_j$. Let $n_x$ be the node that offers the service capability $c_i$ which reaches $n$. Note that during propagation toward $n$, $n_x$ may be replaced if some other node on the path also has the service capability $c_i$. Let $n_y$ and $n_z$ be two nodes on the propagation path from $n_x$ to $n$ and $n_y$ is the immediate neighbor of $n_x$ and $n_z$ is the immediate neighbor of $n_y$.

Now, we define stability of $c_i$ capability in $RT(n_y)$ received from node $n_x$ as:

$$c_i.n_x.stb(n_y,t_i) = overlap(n_x,n_y)$$

where the overlap function is the number of periods that nodes $n_x$ and $n_y$ are within each other's range in the future and $t_i = ct, ct+1, \ldots, ct+k, > ct+k$, and $ct$ is the current time. When an existing $c_i$ is considered for the selection procedure, if $c_i$ is hosted by a mobile node, then $c_i$'s positions may be outdated. In this case, we assume that the mobile node will send its information to its new neighbors and $c_i$ will get to the routing tables of its new neighbors. Thus, we can assume that the current $c_i$ entry is useless after $k$ periods. Then, the $c_i.n_x.stb$ value for $t_i > ct+k$ is 0.

Future positions of $n_y$ and $n_x$ are passed on to $n_z$, because even if $n_y$ is leaving $n_z$ and $n_x$, $n_z$ could be moving toward $n_x$. If we cumulate the future position information of all the nodes on the path, the communication overhead will increase significantly. Thus, we only consider the future positions of 2-hop neighbors. Node $n_y$ will send its stability value with $n_x$ to node $n_z$, then to calculate the stability of $c_i$ in the second hop (in this case $n_z$), we have,

$$c_i.n_y.stb(n_z,t_i) = \max(overlap(n_x,n_z),$$
$$\min(overlap(n_y,n_z), overlap(n_x,n_y))).$$

Considering 2-hop neighbors, if $n$ is the neighbor of $n_z$, then future positions of $n_x$ will disappear in $n$. Even if $n_x$ may move within the range of $n$, we assume that it will happen in the further future, not in the near future. Hence, the stability of $c_i$ in $n$ will be,

$$c_i.n_z.stb(n,t_i) = \max(overlap(n_y,n),$$
$$\min(overlap(n_z,n)), overlap(n_y,n_z)). \quad (6)$$

The stability value be calculated for every tuple of $c_i$ in the routing table and for different neighbors upon receiving the information from neighbors. It shows how much a capability is stable based on neighbor's relative mobility. The node can then use this value to calculate the utility value and decide for the selection and summarization procedure when a new update message arrives.

## 5.5 Utility Calculation

Utility is a parameter based on our four discussed metrics and to show the value of advertised and stored capability in the roputing table. We use an integer in the range of $[0, Maxut]$ to represent the utility. The choice of $Maxut$ may depend on the network size, degree of nodes, mobility pattern and the ontology tree. We choose $Maxut = 32$. For each device owned by node itself or the immediate neighbors which are not able to do routing, the utility value is set to $Maxut$. When a node $n$ receives an advertisement message from a neighbor node about a capability $cpb$ with utility value $ut$, $n$ updates the utility value and stores $cpb$ in its routing table.

Under each capability in routing table is a list of tuples, each specifying a neighbor and the utility ($ut$) value for that neighbor. Consider node $n$ with routing table $RT(n)$. For each neighbor $j$ under capability $RT(n)_i$ $RT(n)$ we use calculated values in (3), (5) and (6) and calculate utility of $NB_j$, $RT(n)_i.NB_j.ut$, by:

$$RT(n)_i.NB_j.ut = w_1(RT(n)_i.NB_j.stb)(RT(n)_i.ug(t)) - w_2 * max_j(c_i.oc(a_j) * (RT(n)_i.NB_j.hop)$$

where $w_1$ and $w_2$ are the weighted sum coefficients. Utility will be included in the advertisement messages and neighbor receive this information upon receiving new capability advertisement. In case the $RT(n)_i$ is an ancestor node that is being considered newly, then $RT(n)_i.NB_j.ut$ is the average utility of the summarized capabilities (their maximom $ut$ value).

Summarization in our approach is not only base on capability relationship in ontology tree and routing table bound value for adaptive summarization in [17], but also considers capabilities utility values and categorize them in different categories. For the capabilities with higher utility values, nodes do the summarization only base on $ONIDs$ and their relationships. IoT nodes do summarization more aggressively when capability is in the second category, the lower utility values. For this category of capabilities and when needed, node $n$ makes summarization subtree larger and summarizes more capabilities in the routing table.

# 6 SEMANTIC-BASED ROUTING PROTOCOL FOR DYNAMIC INTERNET OF THINGS (SRP_DIoT)

We consider a decentralized semantic-based routing protocol in IoT networks. In IP-based routing, IP addresses are used as the indices of the routing table. For semantic based routing, routing is for discovering a capability and, hence, the routing table should be indexed by capabilities. Our routing table contains a list of capabilities, each is specified by an ontology node (ONID). Under each capability (ONID) is a list of tuples, specifying a neighbor and a utility value indicating what is the value of that specific neighbor for that capability. Consider node $n$ with routing table $RT(n)$. $RT(n)_i$ denotes the i-th capability in $RT(n)$. So, $<RT(n)_i.NB_j,\ RT(n)_i.NB_j.ut>$, for all j is the list of tuples associated to $RT(n)_i$, where $RT(n)_i.NB$ is set of neighbors that have advertised $RT(n)_i$ or $RT(n)_i$'s descendants to $n$ and $RT(n)_i.NB_j$ is the j-th neighbor in $RT(n)_i.NB$. $RT(n)_i.NB_j.ut$ shows the utility of $j$th neighbor for $RT(n)_i$. A sample routing table structure is shown in Fig. 4.

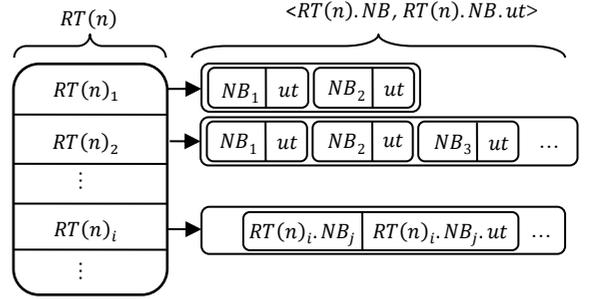

Fig. 4. $n$'s Routing Table ($RT(n)$) Structure

## 6.1 Summarization Algorithm

When node $n$ receives the information about the capabilities of its neighbors, it records the information in its routing table. If $size(RT(n)) > RTB(n)$, then node $n$ summarizes the capabilities in $RT(n)$. As discussed in Section 4, capability summarization is based on the ontology tree done by substituting all the capabilities in a subtree by the capability of the subtree root (in the ontology tree). Let $SL$ denote the level of the subtree potentially to be summarized. We determine $SL$ adaptively to ensure that $n$'s routing table size limit $RTB(n)$ is satisfied. Higher $SL$ implies more aggressive summarization and, hence, lower routing table size and less precise routing.

Summarization not only considers the capability relationship and adaptive summarization but also takes care of capability utility to address mobility of nodes, hop distance and ontology coverage of capabilities and consider their impacts on the summarization reliability. In order to handle this, summarization algorithm first considers $RT(n)_i.NB_j.ut$ and categorize $RT(n)_i.NB_j$ into two categories of high utility and low utility data. Nodes do summarization more aggressively when neighbor is in the second category, the low utility. In this case, $n$ increases the $SL$ value to make subtree larger and summarizes more functionalities in the routing table. The pseudo code for summarizing routing table $RT(n)$, namely, the "routing-table-summarization" procedure, is given in the following.

| function routing-table-summarization ($RT(n)$) |
|---|
| 1    $SL = 0$; |
| 2    $hptRT = \emptyset$; |
| 3      $lptRT = \emptyset$; |
| 4      $sumerizedRT = \emptyset$; |
| 5    **for** each $RT(n)_r$ in $RT(n)$ **do** |
| 6      **for** each $RT(n)_r.NB_j$ in $RT(n)_r.NB$ **do** |
| 7        **if** $RT(n)_r.NB_j.ut \geq utthr$ |
| 8          append $RT(n)_r.NB_j$ to $hutRT$; |
| 9        **else** |
| 10         append $RT(n)_r.NB_j$ to $lutRT$; |
| 11      **endif**; |
| 12      **endfor**; |
| 13    **endfor**; |
| 14    **while** (($size(hutRT) + size(lutRT)) > RTB(n)$) |

```
15      SL = SL + 1;
16      lutRT = summarize-by-level (lutRT, SL);
17      if ((size(hutRT) + size(lutRT))< RTB(n)) then
18        break the while loop;
19      else
20        hutRT = summarize-by-level (hutRT, SL);
21      endif;
22    endwhile;
23    sumerizedRT = hutRT • lptRT;
24    return (sumerizedRT);
25  end function;
```

Function "routing-table-summarization" categorizes available information in the routing table into two different tables of high utility capabilities, $hutRT$, and low utility capabilities $lutRT$ and based on a defined utility threshold $utthr$. Then it first calls function "summarize-by-level" with input $SL$ for low utility data, which returns a new table that is summarized with a maximal summarization level $SL$. If the size of two tables (high and low utility capabilities) is still higher than routing table size limit $RTB(n)$, it calls function "summarize-by-level" with input $SL$ for high utility data too. This way node is summarizing useless information in the routing table more aggressively. The pseudo code for summarize-by-level is given in the following.

```
function summarize-by-level (RT(n), SL)
1   sumRT = ∅;
2   tempRT = RT(n);
3   while tempRT is not empty do
4     RT(n)_k = first entry in tempRT;
5     lsc = ∅;
6     remove RT(n)_k from tempRT;
7     tmptmpRT = tempRT;
8     for each RT(n)_l in tmptmpRT do
9       if RT(n)_l, RT(n)_k are from the same neighbor then
10        remove RT(n)_l from tmptmpRT;
11        <rst, dist_1, dist_2> =
12              common-ancestor (RT(n)_k, RT(n)_l);
13        rst-dist = max (dist_1, dist_2);
14        if (rst-dist ≤ SL)
15          append rst, RT(n)_k and RT(n)_l to lsc;
16          remove RT(n)_l from tempRT;
17          break the inner loop;
18        endif;
19      else
20        remove RT(n)_l from tmptmpRT;
21      endif;
22    endfor;
23    for each RT(n)_l in tmptmpRT do
24      if RT(n)_l, RT(n)_k are from the same neighbor then
25        <newrst, dist_1, dist_2> =
26              common-ancestor (rst, RT(n)_l);
27        dist = max (dist_1, dist_2);
28        if (dist ≤ SL ∧ newrst = rst)
29          append RT(n)_l to lsc;
30          remove RT(n)_l from tempRT;
31        endif;
32      endif;
33    endfor;
34
35    if (‖lsc‖ > 1)
36      calculate rst.ut;
37      append (sumRT, rst);
38    else     //* the node is not to be summarized
39      append (sumRT, RT(n)_k);
40    endif;
41  endwhile;
42  return (sumRT);
43 end function
```

In the pseudo code, $sumRT$ is the summarized routing table to be constructed and returned by the function, $tempRT$ is the cloned routing table $RT(n)$ used as a working space, and summarization is used to store the capabilities that are identified for summarization. The algorithm consists of two nested loops. In the outer while loop, we delist the first capability in $tempRT$, namely, $RT(n)_k$, and compare it with other capabilities, $RT(n)_l$, in $tmptmpRT$ (which is a copy of the current $tempRT$ with $RT(n)_k$ already removed) to find the first capability that can be summarized with $RT(n)_k$ (this is done in the first inner "for each" loop). Capabilities can be summarized if they are similar, i.e., their $onid$s' are within a certain distance, $SL$, to a common ancestor in the ontology tree and are advertised from the same neighbor. Once we find a capability that can be summarized with $RT(n)_k$, the $RT(n)_k$, $RT(n)_l$, and their common ancestor $rst$ are added to the $lsc$ list and $rst$ will be used as the root of the subtree of summarizable nodes (and we exit the first "for each" loop). The second inner "for each" loop finds all the remaining capabilities in $tmptmpRT$ that are descendants within $SL$ levels of $rst$ and puts them in $lsc$ list. All capabilities in $lsc$ list are removed from $tempRT$ list so that they will not be considered further.

If $lsc$ list is empty (i.e., $\|lsc\| \leq 1$) after summarization attempt, which means $RT(n)_k$ cannot be summarized with any capability in $RT(n)$, then $RT(n)_k$ will be added to $sumRT$ alone. If $\|lsc\| > 1$, then all capabilities in $lsc$ will be replaced by the root capability $rst$ (first node in $lsc$) and $rst$ is added to $sumRT$ with the utility value calculated for capability $rst$ (Subsection 5.5).

## 6.2 An Example Scenario

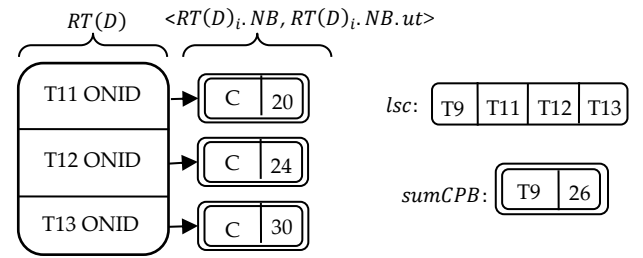

Fig. 5. Routing table and $sumCPB$ construction example.

Consider the example in Fig. 1, which includes a sample ontology tree and a sample network topology with attached devices. Initially, each node adds the ONIDs of its attached devices to its routing table. For example, node $B$ adds <Fixed-LowResolution-DPC(T12), 32> into its routing table. Since each node only has a single capability, so there is no summarization. After the capability information are propagated, $D$'s routing table is shown in Fig. 5.

Next, if D's routing table size is higher than the bound $RTB(D)$, D will try to summarize $RT(D)$ by calling "routing-table-summarization" and subsequently "summarize-by-level". Since T11, T12 and T13 have common parent T9 and are advertised from the same neighbor C, *lsc* list containing T11, T12 and T13 will be generated when $SL$ is 1. Later T9 will replace the other nodes in *lsc* and get inserted into *sumCPB*. T9 will have utility of 26 in this example and after doing utility calculation in $D$. Now, node D can store the *sumCPB* as its routing table, $RT(D)$.

### 6.3 Lookup Algorithm

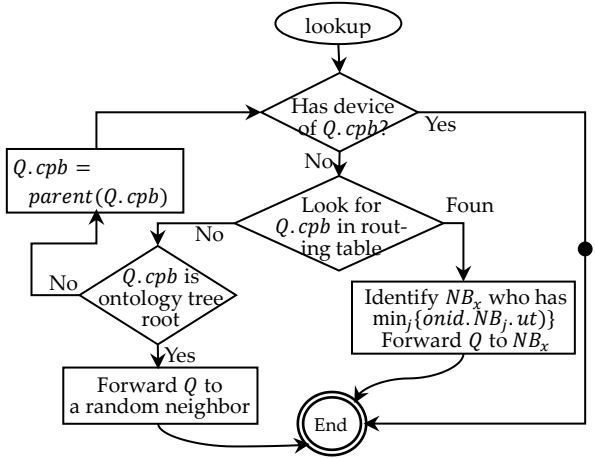

Fig. 6. Flowchart for the look-up algorithm.

Look up is easy when a query $Q$ looking for a capability $Q.cpb$ is submitted or routed to a node $n$. If $Q.cpb$ exists in $n$, routing is done. Otherwise, $n$ searches for $Q.cpb$ in its routing table $RT(n)$ to find $RT(n)_i = Q.cpb$ and then a neighbor with the highest probability to reach the capability. For each neighbor $j$ under capability $RT(n)_i$ in $RT(n)$ we compare utility of $NB_j$, $RT(n)_i.NB_j.ut$. The best neighbor with the maximum utility value will be identified by $n$ among $RT(n)_i.NB$ if ONID of $Q.cpb$ is found in the routing table. Otherwise, $n$ looks for $Q.cpb$ closest ancestor in its routing table and repeats this step till closest ancestor is found in the $RT(n)$. If it reaches the root capability of the ontology tree but still cannot find it in $n$'s routing table, then the query is forwarded to a random neighbor to let that neighbor do the query routing.

If the desired device/capability does not exist in the network during lookup, then termination will be a problem. We use a bound to limit the number of random forwarding to ensure termination in all cases.

## 7 EXPERIMENTAL STUDY

We implement SRP-DIoT to study its performance in dynamic IoT networks. First, we generate sets of IoT capabilities and put each of them in an ontology tree. Ontology tree generation is based on our analysis of real capabilities of IoT devices. Basic capabilities are leaf nodes in the ontology tree while non-leaf nodes are advanced capabilities considered to group some basic or other advanced capabilities together based on their similarities. Five different ontology trees are generated in our experiments. Each node in the ontology tree can be either a leaf node, or a parent node with 2 to 8 child nodes (choosing based on weighted uniform distribution). Size of generated ontology tree for these five ontologies are 186, 435, 810, 1536 and 3086 which basic capabilities in each ontology tree are of size 112, 268, 547, 1190 and 2356. The number of bits required for $ONID$, $\|ONID(n)\|$, for these ontology trees are 32, 32, 48, 64 and 128 bits, respectively.

For the simulated IoT network, we use a mobility generator, Bonnmotion [18], to produce networks of different sizes, from 5000 to 30000 nodes. Nodes in the network are considered to have different mobility patterns and velocities. For each generated topology, we have considered fixed node, medium-speed nodes with average speeds up to 25 m/s, and high-speed nodes with average speeds up to 50 m/s. $R$ percent of the nodes in the network are selected and assigned an IoT capability (ONID of a leaf node in the ontology), and $R = 60\%$ in the experiments. We use Zipf distribution with skew value 0.75 to generate the random capability assignment to nodes. The initial routing table on each node includes the ontology code of its assigned device with the maximum utility calue. We implement SRP-DIoT and other algorithms on ns-3 simulator to compare their performance.

For capability query, we generated queries with different time intervals of 1 second, 30 seconds and 2 minutes. Also, routing table advertisement has been done by 30 seconds, 2 minutes and 5 minutes time intervals. We used different query and advertisement interval pairs as $(Q, A)$ to study different scenarios in performance comparisons. The query capability is also generated following Zipf distribution and the starting node in the network is assigned following a uniform distribution.

We study SRP-DIoT in four different aspects of: **1)** performance comparison with existing semantic based routing protocols, **2)** performance comparison of mobility consideration in hybrid solution compared with existing routing approaches, **3)** impact of mobility consideration in summarization procedure, and **4)** impact of utility concept on the performance of SRP-DIoT. In continue, we show results of experiments for each of these aspects.

**Semantic Routing Algorithms.**

We compare SRP-DIoT with several other representative semantic-based routing algorithms: 1) A *centralized* approach where each IoT node registers its capability to a registry hosted by a central server. The look up query contacts this central node to find the IoT node with the queried capability. 2) The *flooding based* approach. Each node, upon receiving the service discovery query, broadcasts it to all its neighbors. We use a hop limits to control the broadcast. 3) The *caching based* approach, GSD. In GSD each node maintains a cache which caches the historical knowledge about capabilities of the neighboring nodes within a limited number of hops (5 in the experiments). When the cache of a node $n$ is full, $n$ simply flushes out the oldest capabilities to accommodate the newly received capabilities and service groups (timestamps are used for determining the freshness). 4) The *distributed hash tables* (DHT) approach. We map IoT nodes to a Chord ring. Since the IoT devices cannot be moved like the documents, the successor node does not hold the IoT device, but just keeps track of the capabilities hashed to its hosting range. A service discovery query is routed to its destination using the finger table and the destination node returns the list of all IoT nodes providing the capability. From the list, the source node can choose the nearest IoT node to obtain the service.

For both GSD and SRP-DIoT, we set $RTB(n) = 1$ MB for all nodes. Note that the flooding approach does not require routing table and centralized and DHT based approaches cannot have routing table size limit. To ensure a fair comparison, we added backtracking in query lookup for SRP-DIoT and GSD algorithms to ensure 100% success rate. It may not be fair to ensure 100% success rate for the controlled flooding scheme because of its high traffic volume for query routing. Instead, for flooding, we try to set the hop limits for different scenarios so that the success rates are very close to 100% and we record the actual success rates for reference.

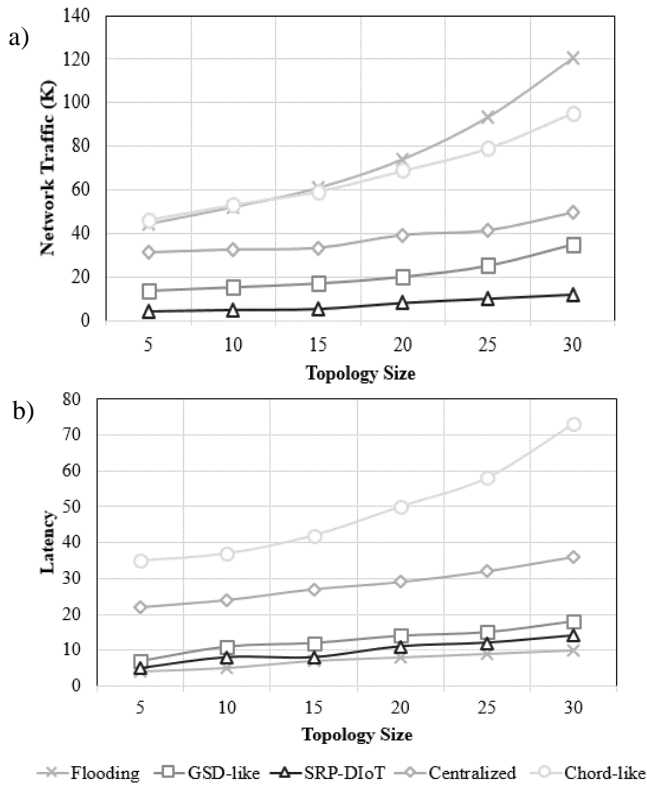

Fig. 7. Performance comparison of various protocols (*Mobility = (20,50,30), (Q,A) = (30,120) and number of capabilities = 1190*)

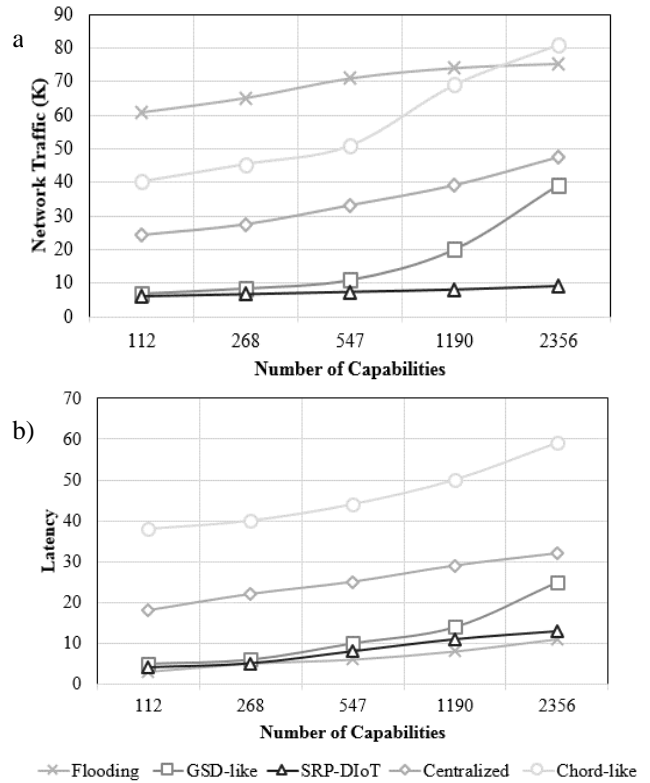

Fig. 8. Performance comparison of various protocols (*Mobility = (20,50,30), (Q,A) = (30,120) and topology size = 20k*)

First, we compare various algorithms with different network sizes. The ontology tree is of size 1190 (number of leaf nodes) and the percentage of mobile nodes are (20,50,30) indicating 20% fixed nodes, 50% medium-speed nodes and 30% high-speed nodes. Queries are generated by 30 seconds time intervals and each node advertises its routing table every 120 seconds, $(Q, A)=(30,120)$. Fig. 7 shows the average traffic volume and the latency for doing queries.

As can be seen from Fig. 7, the traffic volume for query routing in our protocol is much lower than the other algorithms. Performance of GSD is the closest to that of ours, but still the traffic volume for query routing grows more sharply with the increasing network size in GSD than in our protocol. Latency is the number of hops for query routing (including the backtracking hops in SRP-DIoT and GSD). As we know the flooding approach achieves the optimal latency. SRP-DIoT is the second approach as is very close to the optimal latency in all network sizes.

Next, we explore the impact of ontology tree and do performance comparison for different number of available capabilities in the network. We fix network size to 20 k in this comparison and Fig. 8 shows the result of experiments.

Fig. 8 shows that routing table summarization in SRP-DIoT helps this approach to more efficiently handle higher number of capabilities especially compared with the cache based approach, GSD, which has the closest results to SRP-DIoT in smaller ontologies. In continue we will study mobility and our mobility consideration impacts on the query routing.

**Mobility Consideration.**

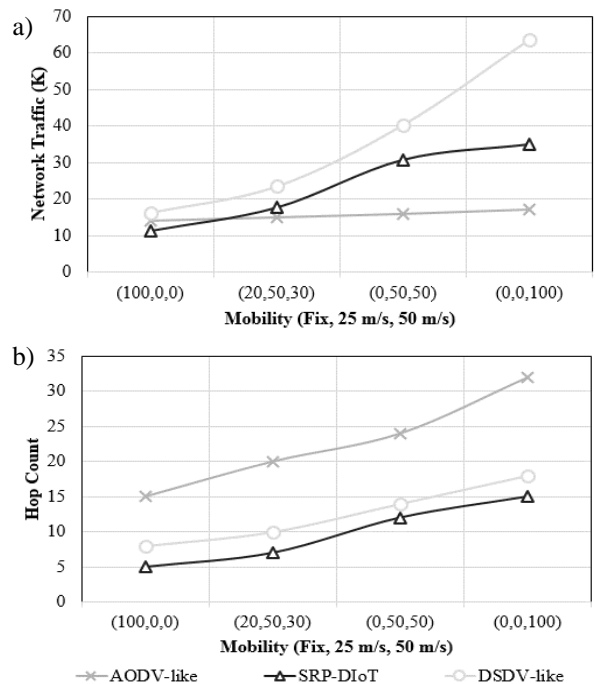

Fig. 9. Performance comparison of various protocols (*Network size = 20k, and number of capabilities = 1190*)

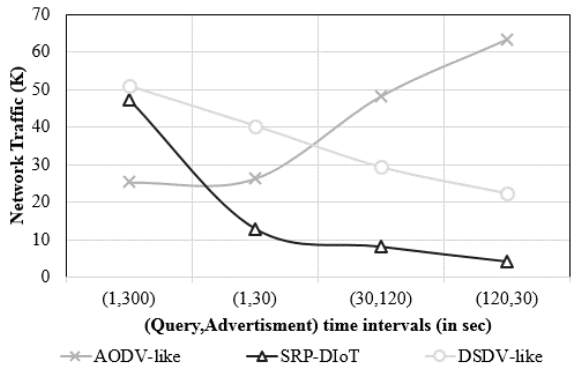

Fig. 10. Network traffic for various approaches (*Network size = 20k, Mobility = (20,50,30) and number of capabilities = 1190)*)

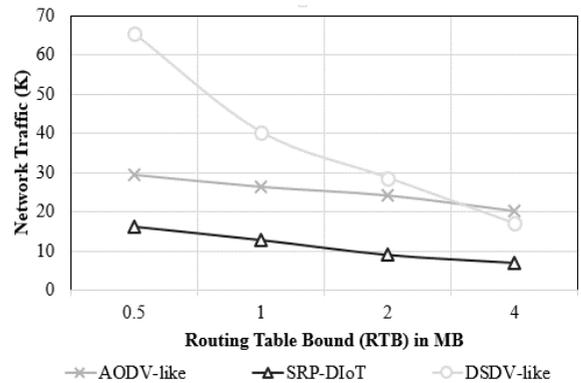

Fig. 11. Network traffic for various approaches (*Network size = 20k, Mobility = (20,50,30) and number of capabilities = 1190)*)

We simulated an AODV-like protocol as on-demand routing protocol and a DSDV-like routing protocol as table-driven approach. For AODV-like protocol, each node cache information in its routing table but we indexed cache and generated RREQ messages by capabilities. For DSDV-like approach, we considered SRP-DIoT without mobility consideration and its summarization method. Each node advertises its routing table to all its neighbors and stores all received information from its neighbors. We count SRP-DIoT a hybrid approach which gets the benefit of all these two solutions. We studied the performance of all these routing approaches for different mobility scenarios. Fig. 4 shows the result of experiments where "Mobility (Fix, 25 m/s, 50 m/s)" shows the percentage of available nodes with different speeds in the network. (100,0,0) means there is no mobile node in the network and all 100% percent of nodes are fixed nodes.

As can be seen, SRP-DIoT is the best among these solutions for stable networks. The AODV-like approach does not need advertisement messages and has better condition in terms of network traffic when mobility is high in the network. However, if cache does not include appropriate information, this approach needs to do flooding upon receiving a query and has higher response time related to table-driven solutions which store capabilities in their routing tables. Fig. 9 also shows that the DSDV-like approach is much more sensitive to the mobility than SRP-DIoT. That's because SRP-DIoT gets advantage from mobility consideration in doing advertisements and constructing routing tables.

Next, we studied the impact of different time intervals for query and advertisement messages generation. We explored this by generating queries and advertisement messages in different scenarios, from (1,300) scenario where queries generated every seconds and nodes advertised their routing information every 5 minutes, to the (120,30) scenario where queries issued rarely compared to the advertisement procedure. Fig. 10 shows result of this exploration.

From the result shown in Fig. 10, we understand that SRP-DIoT is much better when there is lower advertisement interval compared to the query generation rate. It is higher for rarely advertisement scenario because of our backtracking approach and since nodes have not enough information in routing table and are not able to do routing properly. It also shows that, AODV-like approach is better when queries are generated frequently (every 1 second). However, when the query generation is not happenening frequently (120 seconds), cache information is not helpful and this approach needs to do flooding for every single query which makes it very bad.

Next, we also studied the impact of available memory size on each of these approaches. We explored this by doing experiments for different routing table bound values equal to 0.5, 1, 2 and 4 MB. Fig. 11 shows how *RTB* impacts the network traffic in each approach. DSDV-like does not have routing table summarization and results show that it is very sensitive to the size of routing table. As available memory size increase, all of the routing approaches do better while for all of the routing table bound values SRP-IoT has the best results.

**Utility Considerations.**

Last explorations in experimental study are about the impact of mobility and ontology coverage considerations on the SRP-DIoT. To do these experiments we first removed mobility consideration in utility and studied its effects. Then, we removed ontology coverage consideration and studied its impacts.

As can be seen in Fig. 12, while we have the same performance for network with all fixed nodes scenario, but mobility consideration helps SRP-DIoT to have much better performance in high mobility scenarios. Fig. 12 also shows that acncestor coverage helps SRP-DIoT to reduce the network traffic especially in large network topologies. This is also important especially when summarization is coarse grain and summarization level required to be high in order to support routing table bound. This situation happens when we grow the ontology and increase number of available capabilities in the network. Fig. 12 shows how ontology coverge in utility helps SRP-DIoT to reduce network traffic in different sizes of ontology.

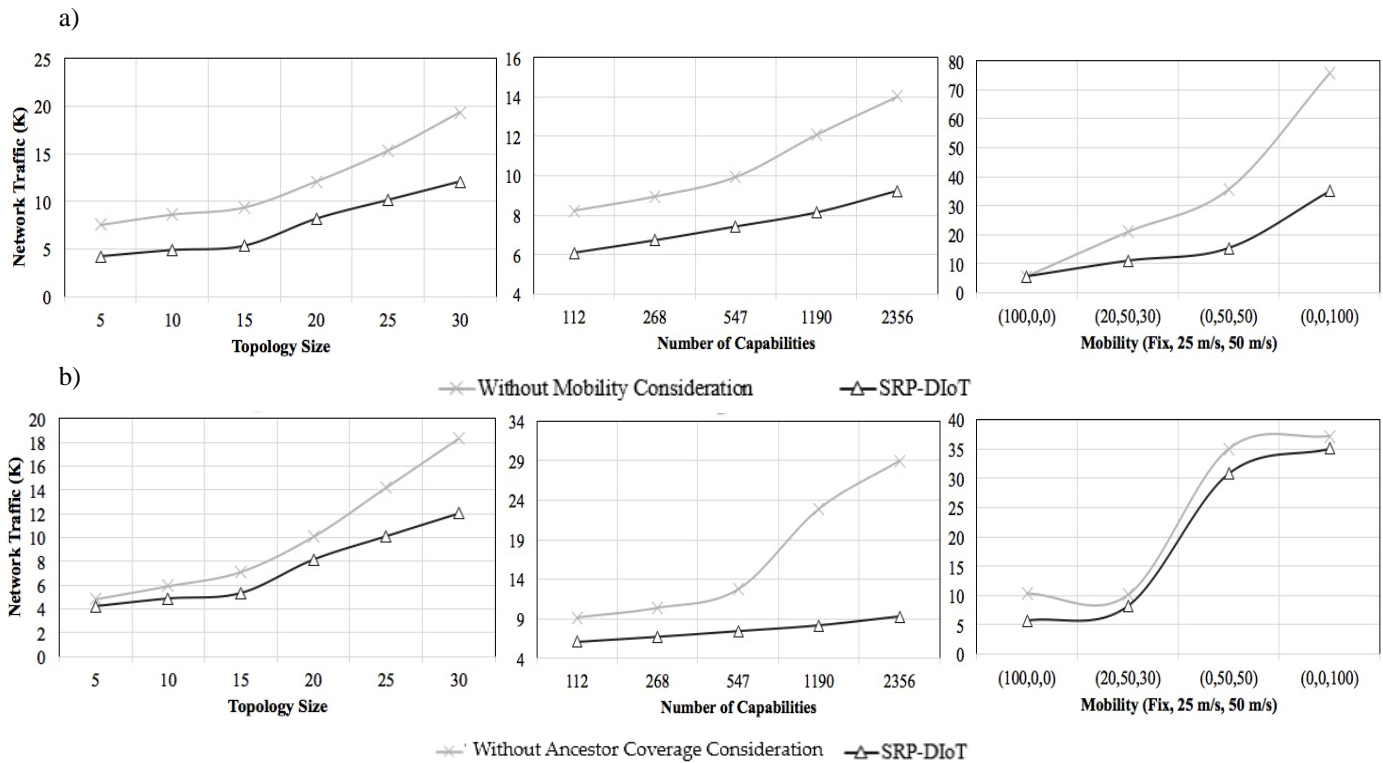

Fig. 12. Network traffic comparisons in SRP-DIoT for (a) mobility consideration, and (b) ontology coverage consideration in summarization algorithm

## 8 Conclusion

We designed and introduced SRP-DIoT, a semantic-based routing protocol for dynamic IoT networks, to efficiently do dynamic arising tasks in dynamic topologies due to IoT devices mobility. To reduce the space requirement for routing, we designed an ontology-based summarization algorithm to smartly group similar capabilities in the routing tables and support adaptive routing table compression. We then introduced an ontology coding scheme to code capabilities in routing tables and messages. Ontology coding also makes routing table summarization possible without the need of storing the whole ontology tree on resource constrained IoT node.

Summarizing capabilities and replacing them with one capability may cause query misleading. We introduced four different metrics in SRP-DIoT which indicates the inclusiveness of replaced capability, mobility pattern, hop distance and the capability usage and is helpful in query lookup procedure.

Our experimental study shows the performance efficiency of SRP-DIoT compared with several other existing solutions including a centralized solution, a DHT-based approach, a controlled flooding scheme, and a cache-based solution.